\documentclass[12pt]{article}
\usepackage{psfig}
\input epsf
\begin{document}

\begin{center}
{\Large Locating where Transient Signals Travel  in Inhomogeneous 
Media\footnote[1]{Portions of this work were presented at the Acoustical Society of America meeting in
New York, NY in May 2004, J. Acoust. Soc. Am. 115, 2549, 2004}  }
\vspace{.2in}

John L. Spiesberger\\

{\it Department of Earth and Environmental Science, 
240 S. 33rd St.\\
U. Pennsylvania, Philadelphia, Pennsylvania 19104-6316}\\
\vspace{.15in}

e-mail: johnsr@sas.upenn.edu\\
\vspace{.15in}

31 January 2005
\end{center}

\begin{center}
ABSTRACT
\end{center}
Locating where transient signals travel between a source and receiver requires
a final step that is needed after using a theory of diffraction such as
the integral theorem of Helmholtz and Kirchhoff.  Introduced here, the final step
accounts for interference between adjacent apertures on a phase screen
by adaptively adjusting their phase and amplitude, yielding a hierarchy of energy  contributions
to any desired window of signal travel time at the receiver. 
The method allows one to check errors in ray theory at finite wavelengths.
Acoustic propagation at long distance in the oceanic waveguide 
(50-100 Hz, 0.05 s resolution) has significant deviations from ray theory.
The boundary condition of zero pressure at the surface of the ocean 
appears to cause sound to travel in a nearly horizontal 
trajectory for a much greater distance near the surface than 
predicted by rays. 
The first Fresnel zone is an inappropriate scale to characterize where transient sounds travel
near a ray path as assumed by a standard scattering theory.  
Instead, the Fresnel zone
is too large by an order of magnitude for cases investigated here. Regions where sounds travel
can have complicated structures defying a simple length scale.
These results are applicable to the physics of underwater sound,  
optics,  radio communication, radar,  geophysics, and theories of wave-scattering.
\vspace{.3in}

PACS numbers: 42.25.Fx, 43.20.El, 43.30.Re\\

\begin{flushleft}
{\bf I. INTRODUCTION}
\end{flushleft} 

Methods that quantify {\it where} transient acoustic and electromagnetic signals propagate between 
a source and receiver in inhomogeneous media have numerous applications.  Acoustic applications
include underwater communication systems and scattering theories from 
fluctuations in the air, the solid  Earth, Sun, and the ocean. 
Electromagnetic applications include the Global Positioning System, radar, and 
communication systems that are affected
by scattering from fluctuations in the index of refraction.
It is important to realize that, except for ray theory, propagation models that
compute the field at a receiver from emissions at a source do not reveal {\it where}
transient signals propagate to reach the receiver. Instead, they show what the received
field is.  Locating regions where transient signals travel at
finite wavelength requires solving Helmholtz's
equation twice, once from the source to receiver, and the other from the receiver to the 
source \cite{born,pierce}.
The integral theorem of Helmholtz and Kirchoff uses these two solutions
to estimate the received signal due to the field passing through any
aperture on a phase screen between the source and receiver at finite wavelengths.
This theory and that for rays
have been
used for more than a century.  The literature on the propagation of sound in the sea
appears to have a single pioneering paper discussing how a diffraction theory based on physical 
optics \cite{born} compares to the theory of 
rays for transient signals at low frequency \cite{bowlineig}.  That paper yields
paths that are ray-like, thus seeming to confirm the ray approximation.  Ref. \cite{bowlineig} uses
a final step  to estimate where sounds travel.    Later, 
we will see
why that final step is not quite what is needed.  
Without some final step following use of a theory of diffraction that utilizes two solutions
to Helmholtz's equation, it may not be possible to estimate {\it where} transient signals
propagate between a source and receiver.  

This paper introduces the needed final step.
It computes where transient signals propagate based on a hierarchy of energy contributions
to any specified window of travel time at a receiver.
The step adaptively adjusts the phase and amplitude on each phase screen to account
for effects of interference of transient signals between adjacent apertures on a screen.

An interesting theory has contributed many ideas to how
internal waves affect fluctuations of transient sounds in the ocean \cite{fdmwz,colosi}.  The present paper checks 
one of the assumptions used in that theory, namely that transient signals
travel within a first Fresnel zone of a ray path.
This assumption was questioned by
the authors of Ref. \cite{colosi}, but an answer appears to have not been published.

Using the needed final step introduced in this paper, 
we find that the Fresnel zone assumption is not valid.
Actual regions where sounds travel can be  complicated and defy representation by a single
length scale.  
Although it is known that ray theory is expected to be less accurate
at low frequencies, accurate quantification of error has apparently not been shown in the literature
for inhomogeneous media.

From an intuitive standpoint, 
the Fresnel radius cannot be the correct scale to use
for transient emissions.  For example, consider propagation in a smooth waveguide
for two signals having the same center frequency
but different pulse resolutions.  Assume the Fresnel radius is evaluated at the center frequency \cite{fdmwz,colosi}.
Signals from a narrow pulse should remain closer to the ray than a wide pulse because
the signal arrives within a narrower time window then do signals from a wide pulse.
Therefore, a 
Fresnel zone cannot account for the behavior of signals with  different bandwidths.

The unsuitability of the Fresnel radius for describing where transient signals
propagate from a source to a receiver in {\it homogeneous} media
was apparently discovered in 1970 in the context of geophysical 
exploration \cite{trorey}-\cite{zavalishin}. Later, these ideas
migrated to optics where the theory was confirmed with 
an experiment \cite{pearce}. 
The next question is whether this transient theory 
can be extrapolated to yield results for inhomogeneous media.  
Unfortunately but not surprisingly, the results in this paper show that
propagation in inhomogeneous media is quite different than the simpler
situation for homogeneous media.  This paper reveals the complexities
of the physics of propagation.  This is an important first step toward
developing  theories to explain the complicated phenomena.

We start by summarizing results from the propagation of transient signals
in homogeneous media. These results are  applicable for some scenarios in the ocean. They also help motivate
methods for quantifying where sounds travel in inhomogeneous media.

\begin{figure}
\psfig{figure=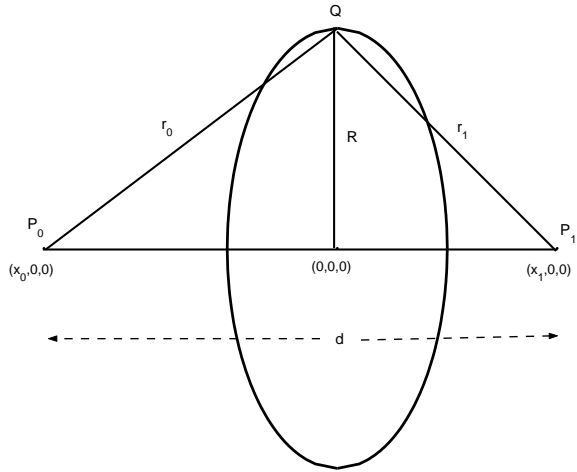,height=2.5in}
\begin{center}
\caption{Geometry for calculation of acoustic field emitted
at $P_0$ and received at $P_1$ at distance $d$ through circular opening of radius $R$.}
\label{f:fig_circ_opening}
\end{center}
\end{figure}

\begin{center}
{\bf II.  HOMOGENEOUS MEDIUM: ZONE OF INFLUENCE FOR TRANSIENT EMISSIONS}
\end{center}

Consider propagation of any transient signal originating from a point source at $P_0=(x_0,0,0)$ to a receiver
$P_1=(x_1,0,0)$ through a circular opening of radius $R$ 
(Fig. \ref{f:fig_circ_opening}).  The opening is perpendicular to the line between source and receiver,
and the spatially homogeneous speed of sound is $c$.  When  the emitted signal is $\alpha(t)$ with $t$ denoting
time, the exact solution of the acoustic wave equation is obtained using the 
general form of Kirchhoff's theorem for transient signals \cite{trorey}-\cite{pearce},
\begin{equation}
{\cal V}(t)= \frac{\alpha (t- d/c ) }{d} - \biggl(\frac{|x_0|}{r_0}+\frac{x_1}{r_1} \biggr) 
\frac{1}{2(r_0+r_1)} \alpha \bigl(t-\frac{r_0+r_1}{c} \bigr) \ , \label{eq:ciraper}
\end{equation}
where $d$ is the distance between source and receiver,
\begin{equation}
d \equiv x_1-x_0 \ . \label{eq:defined}
\end{equation}
Eq. (\ref{eq:ciraper}) has a contribution from the signal along the straight path with geometric
delay $d/c$ plus an inverted diffracted echo traveling along the line segments
$\overline{P_0Q}$ and $\overline{QP_1}$ with geometric-like delays
of $r_0/c$ and $r_1/c$ respectively (Fig. \ref{f:fig_circ_opening}).  Interference occurs
between the geometric and diffracted echo as long as they are separated by less
than the temporal resolution of the signal, $T$. When the radius of the circle exceeds $R=R_I$,
the two contributions are separated in time where the first contribution, $\alpha (t- d/c)/d$,
is that obtained from propagation in a homogeneous medium without screens or boundaries.
Thus, the region of space traversed by propagation in a homogeneous medium without screens 
is,
\begin{equation}
R=R_I(x)=\frac{1}{2}\frac{ \big[cT(2|x|+cT)(2|d-x|+cT)(2d+cT)\big]^{1/2}}{d+cT} \ .
\label{eq:R_I_exact}
\end{equation}
For $cT<<x$ and $cT<<d-x$,
\begin{equation}
R_I(x) \cong \sqrt{\frac{2cTx(d-x)}{d}} \ .  \label{eq:R_I}
\end{equation}
where $x \equiv |x_1|$.
The diffracted echo vanishes as the radius of the circle goes to infinity, but in any case has no effect
on the free-space solution for $R\geq R_I$.  The diffracted component that
reflects from the edge of the opening is consistent with the ideas from the geometrical theory
of diffraction \cite{keller}.

Eq. (\ref{eq:R_I}) has the same form as the Fresnel radius,
\begin{equation}
R_f = \sqrt{\frac{\lambda x (d-x)}{d}} \ . \label{eq:fresnelradius}
\end{equation}
when the pulse resolution, $T$,
is replaced by half the period of the single frequency, $\lambda/2c$ where the monochromatic
wavelength is $\lambda$.  Note, $R_I/R_f$ is $\sqrt{2Q}$ where $Q$ is the quality of the signal
so there is not much difference unless $2Q$ is large.
Significantly, 
Eq. (\ref{eq:R_I_exact}) only depends on the temporal resolution.  It does not depend
on the center frequency as does the Fresnel radius.
$R_I$ is called the ``zone of influence'' \cite{bruhl}.

This paper considers where signals travel from a source that contribute  energy to a received travel time within
one pulse resolution.  This gives rise to the concept of the zone of influence which is the relevant
physical scale of interest in this paper.  This is not the same as the
generalization of the Fresnel radius for transient signals \cite{pearce}.  The generalized
Fresnel radius has a different value than the radius for the zone of influence for homogeneous media. 
Additionally, the generalized Fresnel radius approaches the Fresnel radius as the bandwidth
goes to zero.  However, $R_I$ properly goes to infinity as the bandwidth goes to zero because the pulse resolution
goes to infinity thus admitting sound from an infinite distance for an infinitely long pulse.

\begin{flushleft}
{\bf III. DIFFRACTED REGIONS IN INHOMOGENEOUS MEDIA}
\end{flushleft}

The contribution to the time series at time $t$ at the receiver
from the wave field passing through a transparent opening between depths $z_r$ and
$z_s$ in an otherwise opaque vertical screen at horizontal coordinate $x_{sc}$  is,
\begin{equation}
{\cal G}_j(t,x_{sc},z_r,z_s) = B_j {\cal F} \Big\{ \int_{z_r}^{z_s} \int_{-\infty}^{\infty} I_j(x_{sc},z,\omega) 
\exp (-i \omega t) d\omega dz \Big\}  \ ; \ j=1,2 \label{eq:G} 
\end{equation}
where $B_j$ is a normalization constant and $\omega$ is the radian frequency.
The function ${\cal F}$ removes the carrier frequency
as in Eq. (47) of Ref. \cite{tappert} via complex demodulation of analytic signals.  
The subscript $j$ denotes that diffraction is estimated from the 
integral theorem of Helmholtz-Kirchhoff \cite{born,pierce}  in which case,
\begin{equation}
I_1(x_{sc},z,\omega) = W_1(x_{sc},z,\omega) \frac{\partial W_0(x_{sc},z,\omega)}{\partial x}
- W_0(x_{sc},z,\omega) \frac{\partial W_1(x_{sc},z,\omega)}{\partial x} 
\end{equation}
or from physical optics$^{1}$ in which case,
\begin{equation}
I_2(x_{sc},z,\omega) = W_0(x_{sc},z,\omega) W_1(x_{sc},z,\omega) \ , \label{eq:integrand_geometricdiffraction}
\end{equation}
(Eqs. 18,23 of Ref. \cite{bowlineig}).  The solutions of the Helmholtz equation on the screen 
due to emissions located at the source and receiver are
$W_0(x_{sc},z,\omega)$ and  $W_1(x_{sc},z,\omega)$
respectively \cite{born,pierce}.  In this paper, the inclination factor \cite{born} for physical optics 
is
set to unity because the direction of signal propagation is almost perpendicular to the phase screens.

Let the ``region of diffraction'' denote locations where sounds travel. 
This region was estimated for oceanic propagation \cite{bowlineig} 
by contouring a normalized function of Eq. (\ref{eq:G}),
\begin{equation}
\tilde{{\cal G}}_j(t,x_{sc},z) \equiv \frac{{\cal G}_j(t,x_{sc},z,0)}{{\cal G}_j(t,x_{sc},-D,0)} \ ; \ j=1,2 \ , 
\label{eq:normG} 
\end{equation}
where $D$ is the depth of the bottom of the phase screen at horizontal coordinate $x=x_{sc}$.  The bottom
of the phase screen is chosen such that no significant energy propagates to the receiver below D.
The top of the phase screen is set at the pressure release surface of the ocean, $z=0$.  The function,
$\tilde{{\cal G}}_j(t,x_{sc},z)$, goes from zero at $z=0$ to unity at $z=D$ but not 
necessarily in a monotonic fashion.

Three innovations are introduced that yield accurate estimates of the region of diffraction
based on a hierarchy of energy contributions at the receiver.
Firstly, we consider the diffracted region responsible for the energy
arriving between $t \ \pm T/2$ at the receiver instead of  the region responsible
for the energy at time $t$ as before \cite{bowlineig}.   Secondly,
a method is provided for drawing boundaries of diffracted regions based on a calibration
with results from homogeneous media instead of apparently choosing arbitrary boundaries \cite{bowlineig}.
Thirdly, an adaptive method is introduced for computing the contribution of each aperture
on a phase screen to the energy arriving
within $t \ \pm \ T/2$ at the receiver.
This method properly
accounts for significant effects of interference between apertures on a screen.
Ref. \cite{bowlineig} apparently suppressed ripples of $\tilde{{\cal G}}_j(t,x_{sc},z)$ in Eq. (\ref{eq:normG})
because they were believed to be caused by the interference of separate paths (p. 2667 of \cite{bowlineig}).
We show  that ripples can be due to other phenomenon,
such as self-interference within a resolved path, and should be not suppressed nor are they undesirable.
Instead, the ripples are critical to measure to obtain a proper accounting  of the energy passing
through any aperture on the way to the receiver. 
A demonstration is given
first for propagation in a homogeneous medium.

Assume a source and receiver are placed at 2 km depth in the sea and separated
by 9.129 km over a flat bottom at 5.6 km.  To minimize sidelobes
in the time domain, a Hann taper is applied in the frequency domain
to the emitted signal between $100 \pm 40$ Hz.  
The taper is zero at 60 and 140 Hz, yielding
an effective bandwidth of 40 Hz and a time resolution of $1/40=0.025$ s.
The speed of sound is 1.5 km/s.
The theoretical time for the horizontal path to reach the receiver is 
$9.129/1.5 = 6.086$ s.  The next 
path at 6.645 s corresponds to energy that reflects once from the surface. 
The path that reflects once from the bottom has a travel time
of 7.751 s.  All three paths are temporally resolved with
the pulse resolution of 0.025 s.  

The $c_0$ insensitive parabolic approximation \cite{tappert} 
yields estimates of $W_i(x_{sc},z,\omega)$ ($i$=0,1) and its derivative
for each aperture through Eq. (\ref{eq:G}) for
both $j=1,2$.
This parabolic
approximation  yields accurate travel times, is efficient due to
its split-step algorithm, and obeys reciprocity.  It is important
that reciprocity is obeyed because a proof for reciprocity
is provided by the integral theorem of  Helmholtz and Kirchhoff.
Computational grids in depth and range are 
made sufficiently small to achieve convergence of the solution.

\begin{figure}
\psfig{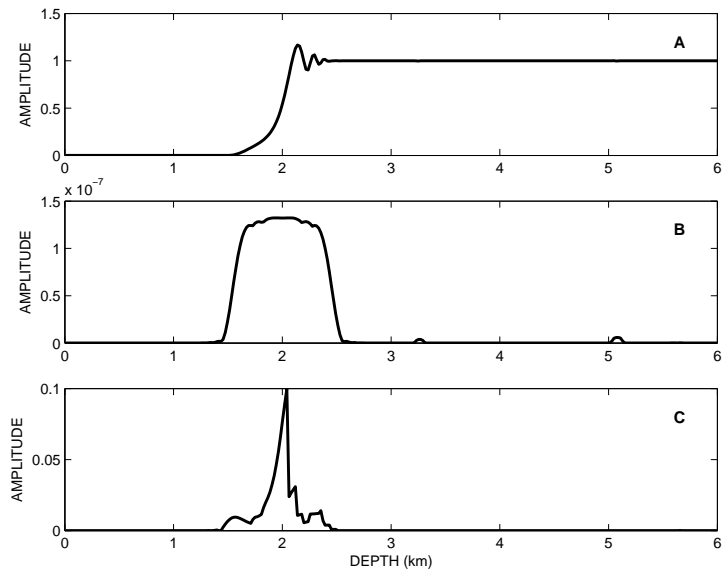}
\begin{center}
\caption{Example of how interference on a phase screen affects calculations
of where sounds travel between a source and receiver in homogeneous media. The 
distance between source
and receiver is 9.123 km and both are at a depth of 2 km. The speed of propagation
is 1.5 km/s. The vertical phase screen is at 9.123/2 km range. Calculations
are for the first (direct) path using a time window with the pulse resolution  of 0.025 s. 
(A) The maximum value of the received time series within the time window 
as a function of depth on the phase screen starting from the surface
and working downward (Eq. \ref{eq:tilderho}).
(B) Contributions to the energy of the received pulse as a function of depth
on the screen
if adjacent apertures did not
interfere destructively at the receiver.
(C). Contributions to energy when interference is accounted for
via Eq. (\ref{eq:deltarho_plus}).}
\label{f:fighomog_screen_contribution}
\end{center}
\end{figure}

Consider the phase screen  half-way between the 
source and receiver (Fig. \ref{f:fighomog_screen_contribution}A).
Cumulative vertical contributions from this screen within times $t \ \pm \ T/2$ can be examined with,
\begin{equation}
\tilde{\rho}_j(z) \equiv \max_{t \in (t \pm T/2 ) } \{\tilde{\cal G}_j(t,x_{sc},z,0)\} \ , 
\label{eq:tilderho}
\end{equation}
which is zero at the surface and unity at the bottom of the phase screen (z=-D).
This function increases monotonically until just after 2 km at which point
interference is observed (Fig. \ref{f:fighomog_screen_contribution}A). 
The apertures on this screen are about 0.02 km wide, so there are about 300 of these
between the surface and the bottom of the screen at 6 km.
The radius of the zone of influence at this location is about 0.4 km (Eq. \ref{eq:R_I}),
so contributions from the screen should die out at depths of $2 \ \pm 0.4$ km, and they do.
If adjacent apertures did not interfere, the contribution at the receiver from aperture $p$ would be
$\tilde{\rho}_j(z_p) - \tilde{\rho}_j(z_{p-1})$ (Fig.  \ref{f:fighomog_screen_contribution}B).

Effects of interference for the homogeneous case are illustrated after describing
the following new algorithm which requires no intervention from an operator to determine
how to deal with interference. 
It provides
high-resolution images of an objective function through a finite difference of cumulative contributions 
of apertures with depth {\it after}
properly dealing with decreasing values of the function with depth, if any.
In this sense, the output of the algorithm is formed in the same way as a probability density
function which is obtained by differentiating a cumulative distribution function.

Starting from the surface, Eq. (\ref{eq:G}) is computed to the bottom of the first
and second aperture, $z_1^+$ and $z_2^+$, respectively. The plus sign denotes downward integration
from the surface.
The contribution at the receiver
is computed for both integrals for the nonlinear objective function of the time series, 
\begin{equation}
\rho_j(0,z^+) \equiv \max_{t \in (t \pm T/2 ) } \{{\cal G}_j(t,x_{sc},z,0)\} \ ; z=z_1^+,z_2^+ \ \ . \label{eq:rho_0toz}
\end{equation}
If $\rho_j(0,z_2^+) \geq \rho_j(0,z_1^+)$, then the we proceed on. Otherwise, destructive
interference is winning so as to decrease the net contribution to the objective function.
The remedy is to consider the contribution to the objective function from both
apertures together instead of separately by replacing their Fourier 
coefficients, $\int_{0}^{z_1^+} I_j(x_{sc},z,\omega) dz$ and
$\int_{z_2^-}^{z_2^+} I_j(x_{sc},z,\omega) dz$, with their average, namely,
$(\int_{0}^{z_1^+} I_j(x_{sc},z,\omega) dz +
\int_{z_2^-}^{z_2^+} I_j(x_{sc},z,\omega) dz ) /2$.  Replacing the separate values by their
average does not change the sum of the contributions from apertures 1 and 2 to either
the time series at the receiver or to the objective function because the integral
from 0 to $z_2^+$ is unaltered.

The next step compares $\rho_j(0,z_3^+)$ with the newest value of $\rho_j(0,z_2^+)$. If 
$\rho_j(0,z_3^+) \geq \rho_j(0,z_2^+)$, then we continue on-wards. Otherwise, destructive interference
is reducing the contribution to the objective function
at the receiver.  The remedy is similar to before. Fourier coefficients from apertures
two and three are replaced by their mean, and this mean is retained unless
$\rho_j(0,z_2^+) \leq \rho_j(0,z_1^+)$ where the last inequality is based on the newly replaced
mean value of the Fourier coefficients (it is necessary to go back and check this
because the Fourier coefficients for aperture two just got modified and one wants to make sure
that the jump from aperture 1 to apertures 1 plus 2 still gives a non-decreasing 
contribution at the receiver).  
If this last
condition is true (destructive interference is winning), 
then it is necessary to replace the Fourier coefficients
of apertures 1, 2, and 3 by the mean of their newest values, which guarantees that contributions
from each cumulative aperture going from 1 through 3 yields non-decreasing values of $\rho_j(0,z^+)$.  This
procedure continues until the bottom of the phase screen is reached.
The contribution to the objective function 
at the receiver from each aperture $q$ is provided by the finite difference,
\begin{equation}
\delta \rho_j(q)^+ \equiv \rho_j(0,z_{q}^+) - \rho_j(0,z_{q-1}^+) \ ; \ q=2,3,4, \cdots P \ , \label{eq:deltarho_plus}
\end{equation}
and for $q=1$, $\delta \rho_j(1)^+ \equiv \rho_j(0,z_{1}^+)$.
These values are guaranteed to be non-negative as desired.  Fig. (\ref{f:fighomog_screen_contribution}C) shows that
the greatest contributions to the peak of the time series occur over a narrower depth range than
obtained if one assumes that the apertures do not interfere (Fig. \ref{f:fighomog_screen_contribution}B).
Panel C displays a slight bias in the center location of the peak which should ideally
occur at a depth of 2 km.  The bias is suppressed by conducting the procedure
leading to Eq. (\ref{eq:deltarho_plus}) from the bottom of the phase 
screen to the surface, yielding values of $\delta \rho_j(q)^-$, and then averaging from both directions,
\begin{equation}
\delta \rho_j(q) = [\delta \rho_j(q)^+ + \delta \rho_j(q)^-)]/2 \ . \label{eq:deltarho_avg}
\end{equation}

\begin{figure}
\psfig{figure=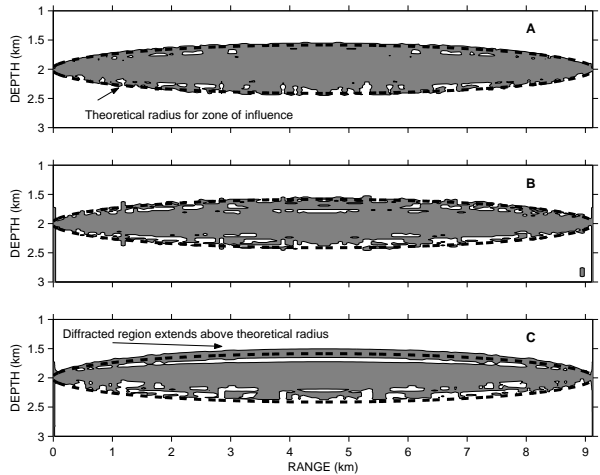,height=2.5in}
\begin{center}
\caption{Three methods for estimating a region of diffraction in a homogeneous medium 
using the integral theorem of Helmholtz and Kirchhoff.  Source and receiver are at a depth of 2 km.
(A).  The relative dB method uses
the top $X_{dB}= -23$ dB contributing apertures on each phase screen based on the
values in Eq. (\ref{eq:deltarho_avg}).
(B). Same except the contributions from each aperture are sorted in non-increasing order
and used until a fraction of  $f=0.90$ of maximum amplitude of the time
series is reached (fractional amplitude method).
(C) Same as (A) except the spatial 
bias is not suppressed because the
phase screen is only smoothed from the top down (Eq. \ref{eq:deltarho_plus}).  Note the region
of diffraction extends slightly above the theoretical radius of the zone of influence (Eq. \ref{eq:R_I_exact})
for a homogeneous medium.}
\label{f:figsize_levels}
\end{center}
\end{figure}

Two methods are described for estimating a diffracted region from Eq. (\ref{eq:deltarho_avg}).
The ``relative dB'' method uses all apertures within $X_{dB}$ of the value of the
maximum contributing aperture on a phase screen.  The ``fractional amplitude'' method sorts 
aperture contributions in non-increasing order, then selects the minimum number of these sorted
contributions from largest to smallest until reaching a specified fraction, $f$, of the maximum amplitude
of the peak of the time series at the receiver. Values for $X_{dB}=-23$ dB and $f=0.9$ 
fit the theoretical value of the zone of influence in a homogeneous medium fairly well (Fig.
\ref{f:figsize_levels}).   The size of the diffracted region is not sensitive to
the values chosen for $X$ and $f$.  The spatial bias of the diffracted region is suppressed when values
of $\delta \rho_j(q)$ are used (Fig. \ref{f:figsize_levels}A) instead of $\delta \rho_j(q)^+$
(Fig. \ref{f:figsize_levels}C).  The diffracted region has holes that indicate 
non-monotonic variations in
sonic level within the zone of influence (Fig. \ref{f:fighomog_screen_contribution}C), 
indicating interference between sound traveling
along the direct path and sound traveling along a diffracted path near the periphery and back
to the receiver.  Therefore, effects of diffracted echoes are evident
even in the absence of an edge on a circular opening ({\bf Sec. II}).

\begin{figure}
\psfig{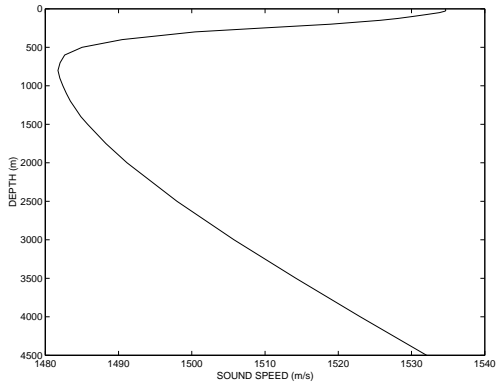}
\begin{center}
\caption{The speed of sound as a function of depth near 
Kaneohe Bay, Oahu (21.512 $^{\circ}$N
and 202.228 $^{\circ}$E) according to climatology \cite{levitus} for Spring
and using Del Grosso's algorithm for the speed of sound in seawater \cite{delgrosso}.
Straight lines are drawn between data points here, but the calculations 
use a quadratic spline \cite{zray} that has no discontinuities in gradient with depth.}
\label{f:fig_cz_kaneohe}
\end{center}
\end{figure}

\begin{flushleft}
{\bf IV. REGIONS OF DIFFRACTION FOR WAVEGUIDE PROPAGATION IN THE OCEAN}
\end{flushleft}

Consider climatological conditions for 
Spring \cite{levitus}  for the North Pacific near  
Kaneohe Bay, Oahu (21.512 $^{\circ}$N
and 202.228 $^{\circ}$E).  
Sound speeds, which only vary with depth, are estimated from temperature, salinity,
and depth using Del Grosso's algorithm \cite{delgrosso} (Fig. \ref{f:fig_cz_kaneohe}). The acoustic models
described below propagate the acoustic field in Cartesian coordinates after
applying the Earth flattening transformation to the field in Fig. (\ref{f:fig_cz_kaneohe}).

Four cases are considered.  The purpose of the first three is to see if
the region of diffraction becomes more ray-like with increasing frequency.
The emitted signal has
a bandwidth of 20 Hz and the center frequency is set at 50, 100, and 200 Hz.
The equivalent pulse resolution is given by the inverse bandwidth, 0.05 s.  
 
The purpose of the fourth case is to see how the region of diffraction decreases
with decreasing pulse width.
The  case repeats the
100 Hz center frequency emission, but the bandwidth is increased to 40 Hz, yielding
a narrower resolution of 0.025 s.

Assume the source and receiver are separated by 523.845 km and placed 
at 800 m depth, corresponding to the minimum speed (Fig. \ref{f:fig_cz_kaneohe}).
The bottom is at 4500 m.  As before, the signal is tapered with a Hann filter
in the frequency domain
to suppress sidelobes in the time domain.  Attenuation of sound by seawater
is accounted for as a function of frequency.

The sound-speed insensitive
parabolic approximation \cite{tappert} incorporates a geoacoustic 
bottom that is specified as follows.
The sediment thickness is 200 m.
The ratio of sound speed  and density at the top of the sediment layer to that
at the bottom of the water column are 1.02 and 1.7 respectively.
The attenuation in the sediment is $\alpha(f) = 0.02 f \ (\mbox{dB/m})$
where $f$ is acoustic frequency in kHz.
The speed increases with depth at a rate of 1 $s^{-1}$ in the sediment.
The ratio of sound speed  and density at the top of the basement to that
at the bottom of the sediment are 2 and 2.5 respectively.  
The attenuation in the basement is $\alpha(f) = 0.5 f^{0.1} \ (\mbox{dB/m})$.

The  software for the sound-speed insensitive approximation has been used before
to identify broadband signals over basin-scales \cite{spiesberger}.  It yields travel times
that agree with exact solutions from normal modes within a few milliseconds
over basin-scales for speeds that vary with depth only \cite{tappert}.

Regions of diffraction are compared
with rays.  The ray program, zray, and its eigenray finder have been described 
and successfully used to study acoustic propagation for many 
experiments \cite{spiesberger,zray}.  Its results agree with analytical 
solutions.  Sound speeds used by the ray program are the same as
those used by the parabolic approximation on its computational grid.
Between grid points, the speed is obtained using a quadratic spline.
The spline goes through each grid point, and has no gradient discontinuities \cite{zray}.

\begin{figure}
\psfig{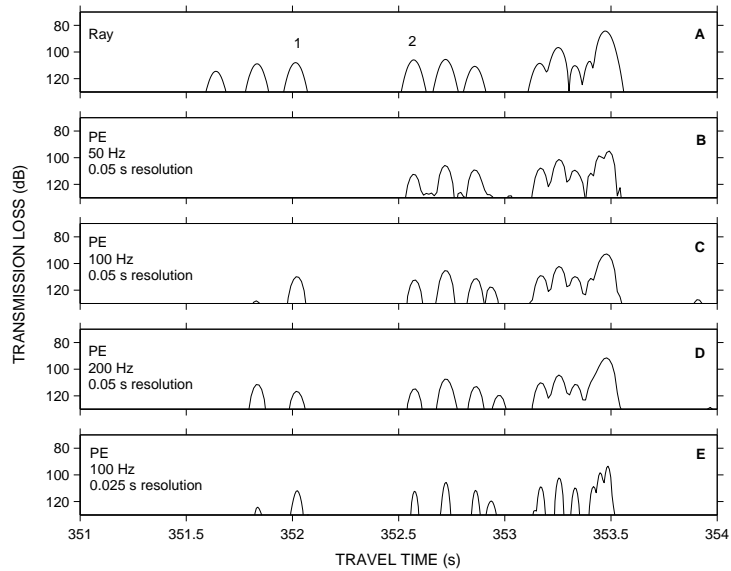}
\begin{center}
\caption{Impulse responses from five models between a source and receiver
separated by 523.845 km and depths of 800 m with peaks 1 and 2 indicated.
The top panel comes from tracing rays through the profile in Fig. (\ref{f:fig_cz_kaneohe}).
The next four panels are obtained using the sound-speed insensitive parabolic 
approximation \cite{tappert} (PE) for the same field of speed as the rays.
The center frequencies for the PE runs are 50, 100, and 200 Hz (panels B, C, D)
with a pulse resolution of 0.05 s. Panel E has a center frequency of 100 Hz
and a pulse resolution of 0.025 s.}
\label{f:fig_newimpulse_c}
\end{center}
\end{figure}

Attention is focused on two peaks that are temporally resolved
at 0.05 s resolution (Fig. \ref{f:fig_newimpulse_c}) (Table I). Each peak is due to a single
resolved ray.   These peaks appear
in all four cases except at 50 Hz where peak 1  is not present 
because of the interaction of low frequency sound with the sub-bottom 
(Fig. \ref{f:fig_newimpulse_c}B).

\begin{center}
\begin{tabular}{ccccc}
\multicolumn{1}{c}{PEAK}&
 \multicolumn{1}{c}{TRAVEL TIME (s)}&
  \multicolumn{1}{c}{\# Turns}&
   \multicolumn{1}{c}{ANGLE (DEG.)}&
     \multicolumn{1}{c}{SHALLOWEST DEPTH (M)}\\
\hline 

1&352.022&19&14.093&130\\
2&352.577&19&-12.577&214\\
\hline
\multicolumn{5}{p{5.5in}}{
{\bf Table I.} Ray parameters for the two peaks shown in 
Fig. (\ref{f:fig_newimpulse_c}).
The angle is that with respect to the horizontal at the source depth of 800 m.
The shallowest ray depth is indicated.}
\end{tabular}
\end{center}

\begin{figure}
\psfig{figure=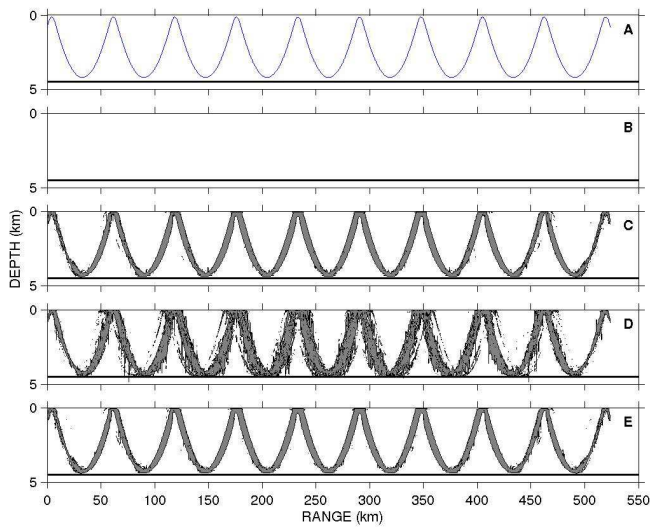,height=3.in}
\begin{center}
\caption{Paths for peak 1 for the same cases in Fig. (\ref{f:fig_newimpulse_c}) 
respectively.
The top panel is the ray, and the others from the
integral theorem of Helmholtz and Kirchhoff (Eq. \ref{eq:G}, $j=1$) using the relative dB method
and Eq. (\ref{eq:deltarho_avg}) to estimate the energy contributions from
each 0.0195 wide vertical aperture on each phase screen. The depth of the bottom at
4.5 km is indicated.
Panels B-D are for center frequencies of 50, 100, 200 Hz respectively
with pulse resolution of 0.05 s.  Panel E is for a center frequency
of 100 Hz and a pulse resolution of 0.025 s. 
Peak 1 does not occur at 50 Hz due to interaction of sound with the bottom (Fig. \ref{f:fig_newimpulse_c}).}
\label{f:peak2_smooth_reldB}
\end{center}
\end{figure}

\begin{figure}
\psfig{figure=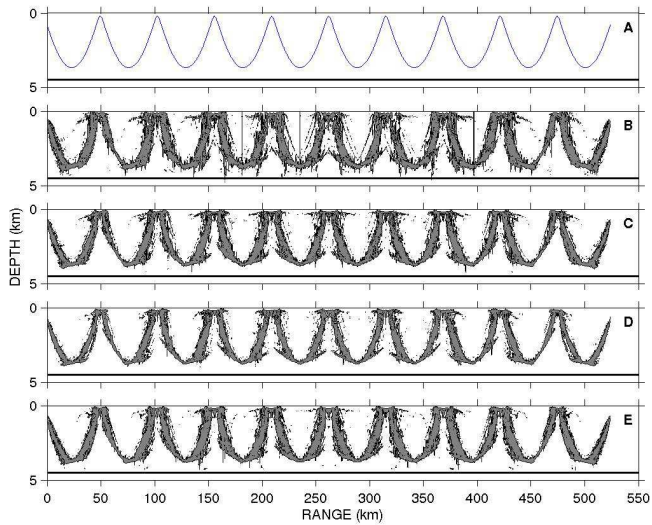,height=3.in}
\begin{center}
\caption{Same as Fig. \ref{f:peak2_smooth_reldB} except for peak 2.}
\label{f:peak3_smooth_reldB}
\end{center}
\end{figure}

The integral theorem of Helmholtz and Kirchhoff is used to compute the results
in this section via Eq. (\ref{eq:G}).  However, results
based on physical optics look almost identical (not shown).
Using the relative dB method that accounts for interference on phase screens (Eq. \ref{eq:deltarho_avg})
diffracted regions
for the  peaks look somewhat like their ray paths with significant discrepancies in many cases 
(Figs. \ref{f:peak2_smooth_reldB}-\ref{f:peak3_smooth_reldB}).
The vertical stripes occurring at some ranges indicate that a wide range
of depths contribute to the relative dB method at the -23 dB threshold on some phase screens.
All diffracted regions show that sound extends over a flatter and more extended
horizontal region near the surface than their ray counterparts (e.g. 40 km instead of O(1) km
for peak 2 in Fig. \ref{f:peak3_smooth_reldB}).
Peak 1 (Fig. \ref{f:peak2_smooth_reldB}C,D) looks less ray like at 200 Hz than 100 Hz 
(Fig. \ref{f:peak2_smooth_reldB}).  Peak 2's upper turning
regions look like they have antlers that extend the exposure to the surface region, particularly
at a center frequency of 50 Hz (Fig. \ref{f:peak3_smooth_reldB}B). The bottoms of the
diffracted regions appear to have attachments of broadly shaped ``V's'' that are more pronounced
at mid range than near the source or receiver.  These ``V's'' are suppressed when the pulse
resolution is changed from 0.05 to 0.025 s (Fig. \ref{f:peak3_smooth_reldB}E).

\begin{figure}
\psfig{figure=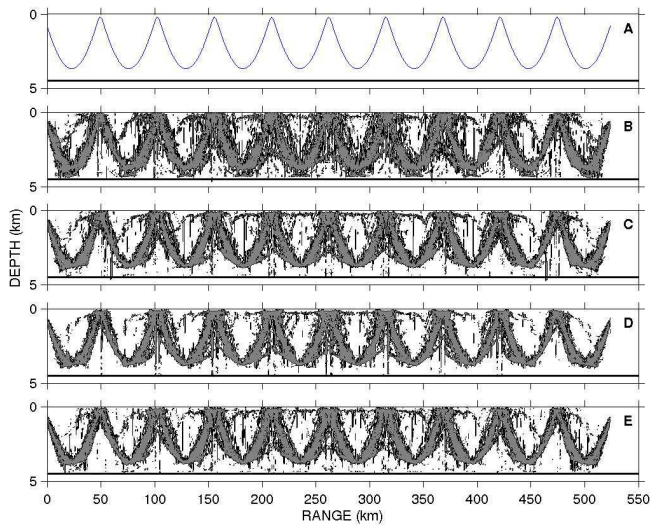,height=3.in}
\begin{center}
\vspace{.1in}
\caption{Same as Fig. \ref{f:peak3_smooth_reldB} for peak 2 except this uses the fractional amplitude method
with $f=0.9$. These diffracted regions are fuzzier and somewhat wider than obtained using
the relative dB method.}
\label{f:peak3_smooth_fracamp}
\end{center}
\end{figure}

The fractional amplitude method for drawing diffracted regions from Eq. (\ref{eq:deltarho_avg})
yields diffracted regions that look a little fuzzier and fatter for peak 2 than those using the
relative dB method (Figs. \ref{f:peak3_smooth_reldB},\ref{f:peak3_smooth_fracamp}).  
Results are similar for the other peaks (not shown).

\begin{figure}
\psfig{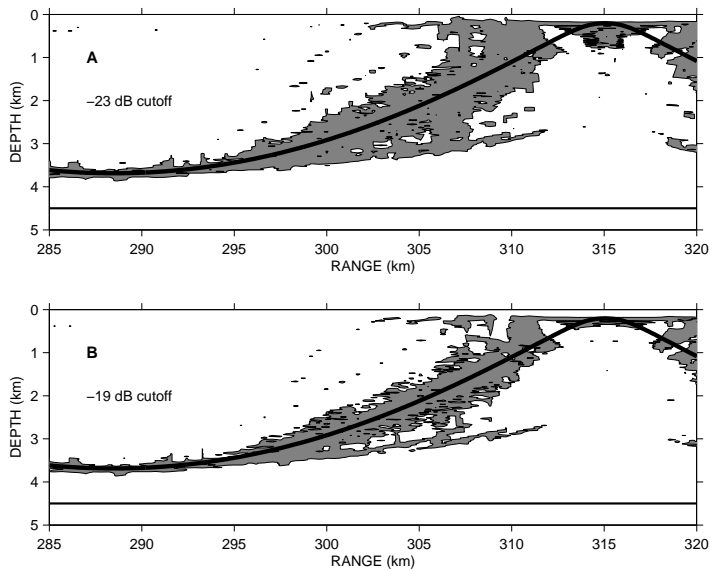}
\begin{center}
\caption{A: Same as Fig. (\ref{f:peak3_smooth_reldB}C) between ranges of 285 and 320 km
with ray path superimposed using a cutoff of $X_{dB}=-23$ dB.  B: Same as A except the cutoff is increased
to -19 dB. The center frequency is 100 Hz.}
\label{f:fig_peak3_blowup_reldb}
\end{center}
\end{figure}

Blowups of the diffracted regions associated with peak 2 near 100 Hz
(Fig. \ref{f:fig_peak3_blowup_reldb}A) allow better scrutiny
of the departures from ray theory.  The threshold criteria
for the relative dB method is changed to suppress some of the
apertures that contribute quieter sounds 
(Fig. \ref{f:fig_peak3_blowup_reldb}B). 
Further suppression
of quieter apertures makes the diffracted regions look discontinuous in space which seems unphysical
because there would not be any path joining the source and receiver.  Thus the 
flat regions near the surface that look
so unlike rays, and some of the other extensions away from the ray path are significant departures 
from ray theory.

\begin{figure}
\psfig{figure=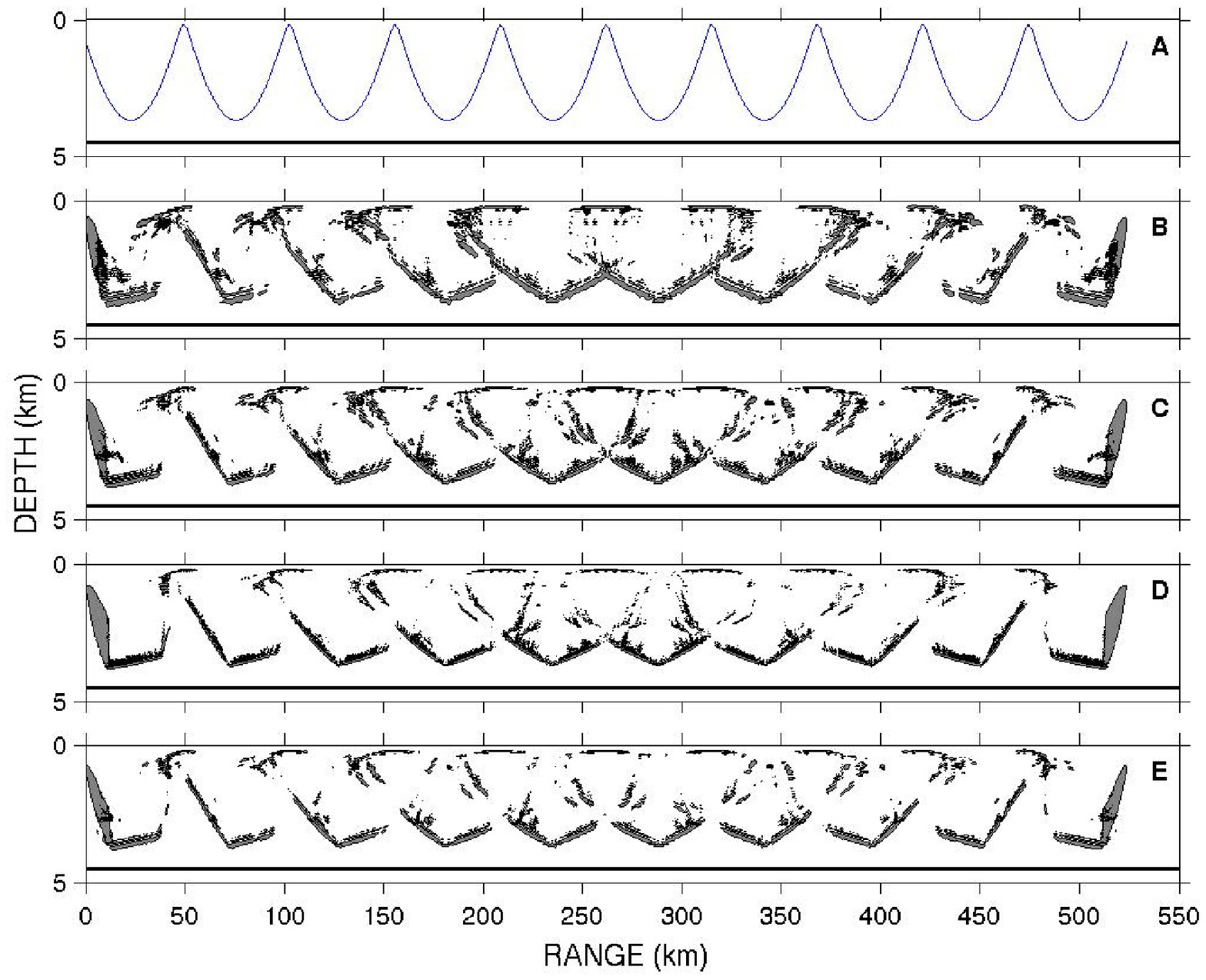,height=3.in}
\begin{center}
\caption{Same as Fig. (\ref{f:peak3_smooth_reldB}) except the effects of destructive interference
between adjacent vertical apertures on phase screens are not 
corrected using Eq. (\ref{eq:deltarho_avg}). Instead this estimate of the diffracted regions
assumes that adjacent apertures do not interfere with each other, yielding an incorrect
picture for where sounds travel. The apertures
within 6 dB of the most energetic aperture on each phase screen are shown  because when interference is
not accounted for, the 6 dB threshold yields the theoretical
zone of influence for homogeneous media (not shown).}
\label{f:peak3}
\end{center}
\end{figure}

\begin{flushleft}
{\bf A. Importance of accounting for interference on a phase screen in inhomogeneous media}
\end{flushleft}

When  the effects of interference are not dealt with using  Eq. (\ref{eq:deltarho_avg}), 
the region of diffraction
looks less plausible than otherwise (e.g. peak 2, Figs. \ref{f:peak3_smooth_reldB},\ref{f:peak3}).
The difference is due to the fact that some adjacent apertures on phase screens pass large
amounts of energy that destructively interfere at the receiver leaving little net energy.

\begin{flushleft}
{\bf B. Comparison with Fresnel radius for inhomogeneous medium}
\end{flushleft}

The first Fresnel radius for any point A on a ray path between a source and receiver 
is the distance, perpendicular to the
ray at A, to a point B such that the difference of the signal wavelength is one-half
between the ray and the perturbed ray passing through B and back to the receiver.
For steeply cycling rays in the deep ocean waveguide such as shown in  Figs. (\ref{f:peak2_smooth_reldB}-
\ref{f:peak3_smooth_reldB}), first order analysis and more exact numerical evaluations
demonstrate the standard result  that the Fresnel
radius for such rays is zero at its turning points and is approximately 
the Fresnel radius for an 
homogeneous medium in-between, i.e. Eq. (\ref{eq:fresnelradius}), 
(see Eq. 7.1.11 in Ref. \cite{fdmwz} and Eq. (26)
in Ref. \cite{colosi}). 
For $d=523.845$ km, $\lambda=1.5 \ \mbox{km} \ \mbox{s}^{-1} / 100 \ \mbox{Hz} = 0.015$ km,
and $x=d/2$ (half-way between source and receiver), the Fresnel radius is $R_f=1.4$ km.
If the Fresnel radius was a good predictor of the region of diffraction, the diffracted 
region would encompass $\pm 1.4$ km on either side of the unperturbed ray path
near 250 km and go to zero at the turning points of the rays in Figs.
(\ref{f:peak2_smooth_reldB}-\ref{f:peak3_smooth_reldB}).  However, the diffracted region for 
peaks 1 and 2 extend for
a length scale of O(0.1) km in a direction perpendicular to the unperturbed rays, and
a radius of 1.4 km is too large for all the diffracted regions shown. The diffracted region
does not go to zero at the turning points.

Replacing the radius of the zone of influence, $R_I$, for $R_f$ does not help this discrepancy
because the length scale 
$2cT=2 \times 1.5 \ \mbox{km} \ \mbox{s}^{-1} \times \ 0.05 \ \mbox{s} = 0.15$ km needed by $R_I$ 
exceeds $\lambda=0.015$ km for $R_f$ at 100 Hz, making $R_I$ about 4.5 km. It is unfortunate but
not surprising that neither the Fresnel radius nor the zone of influence in Eq. (\ref{eq:R_I})
provide the correct scaling for diffracted regions in inhomogeneous media.

\begin{flushleft}
{\bf V. VERIFYING ALGORITHMS}
\end{flushleft}

It is important to verify the correctness of the algorithms because of the fundamental
importance of the results.  The strongest 
evidence for the 
validity of these results 
should and must come from other independent investigations.  For example,
a previous construction of diffracted regions \cite{bowlineig} shows that sound
lingers longer near the surface than the corresponding ray. This behavior  
appears in Figs. (\ref{f:peak2_smooth_reldB}-
\ref{f:peak3_smooth_reldB}) in a much more pronounced way. There are four other
reasons to believe the algorithms are correct.

Firstly, the algorithms  for the inhomogeneous case
are identical to those  for the homogeneous case which agrees with the analytical solution
for the  zone of influence (Fig. \ref{f:figsize_levels}). 
Secondly, the derivation of the integral theorem of Helmholtz and Kirchhoff includes a proof for
the reciprocity of the field in a time-independent medium.   The $c_0$ insensitive parabolic 
approximation in this paper, must also obey reciprocity, and it does \cite{tappert}.  
Thirdly , the time series at the receiver should look like the time series computed
from the integral theorem of Helmholtz and Kirchhoff after integrating over all depths
for any vertical screen between the source and receiver, and they do. 
Fourthly, results from physical optics \cite{born} look
nearly identical to those from the integral theorem of Helmholtz and Kirchhoff for all cases 
in this paper (not shown) 
The similarity is due to the fact that sound propagates nearly perpendicular
to the vertical phase screens, thus the inclination factor from physical optics
should be very near unity as chosen for Eq. (\ref{eq:integrand_geometricdiffraction}).

\begin{flushleft}
{\bf VI. CONCLUSIONS}
\end{flushleft}

A needed last step was added to the integral theorem of Helmholtz and Kirchoff
for the purpose of quantifying hierarchical contributions of energy to any specified
time window at a receiver from transient emissions from a source.  The resulting
construction of regions where signals travel appear to be the first of its
kind for application in inhomogeneous media.  The method introduced here is a fundamental tool
in understanding the physics of transient propagation between two points, and
is useful for quantifying errors in the theory of rays.

Using the new method, the 
first Fresnel radius is shown to be an incorrect scale for estimating where
transient signals travel between a source and receiver in inhomogeneous media.
A scattering theory of sound in the ocean  
adopts the Fresnel radius for determining which fluctuations
affect transient propagation near a ray \cite{fdmwz,colosi}. A re-interpretation of this theory
in light of the findings here should be considered.
This paper finds that the regions where signals travel can
be complicated, and may defy a simple formula for explaining
the physics, such as the Fresnel radius does for single-frequency signals.

A theory that predicts the scattering of sound by internal waves has been shown
to not fit observations for an experiment at 75 Hz center frequency and 35 Hz bandwidth
over 3250 km in the Pacific \cite{colosi}.  The data are consistent with
scattering in the unsaturated or partially unsaturated regime. 
The theory predicts that scattering should be in the fully saturated regime.  The theory would
be more consistent with the data 
if its so-called size parameter, $\Lambda \sim (R/L)^2$ was
several orders of magnitude smaller than its predicted values from 1 to 40 (Fig. 19, Ref. \cite{colosi}).
Here, $R$ represents the
distance perpendicular to an unperturbed ray over which the sound is influenced
by fluctuations, and $L$ is the spatial scale of the fluctuations.  Standard
theoretical and numerical results from that theory 
yield $R =0$ at the upper turning points of the rays and $R \sim O(R_f)$ 
in between which is about 2 km for the 3250 km transmission
(Fig. 20, Ref. \cite{colosi}).  For the simulations discussed in {\bf Sec. IVB}, the diffracted regions
are an order of magnitude less than the Fresnel radius. If the same was true
for the Pacific experiment, $\Lambda$ would decrease by two orders of magnitude. This would
close the gap between theory and observation.  The only point of this back-of-the-envelope
calculation is to raise the possibility that an accurate calculation of the diffracted
region in the Pacific experiment might contribute to a more accurate prediction.

It is important to quantify errors in the ray approximation.
At 100 Hz, the wavelength in the sea is 15 m. This {\it is} small compared
with the 1000 m scale of the wave guide (Fig. \ref{f:fig_cz_kaneohe}).  
The results given here suggest
this ratio needs to be much smaller to consistently yield ray-like paths.
Perhaps the break down of ray theory at low frequencies 
is due to the boundary conditions at the surface
and bottom.  It is also
possible that the pulse resolution needs to be much smaller than 0.025 s to yield ray-like paths
at low frequencies.  Of course, there is a practical limit for the pulse resolution
at low frequency.  The range of validity for Snell's law is apparently 
not well understood at this time, despite
its use for describing sound transmission in the deep sea since
1917 \cite{lichte}. The advance of computers makes it possible to research
this subject in new ways.

\begin{flushleft}
{\bf ACKNOWLEDGMENTS}
\end{flushleft}

This research was supported by the Office of Naval Research contract
N00014-03-C-0155 and by a grant of computer time from the
DOD High Performance Computing Modernization Program at the Naval Oceanographic Office.
The zray raytrace program was modified from
the program written by James Bowlin.  I thank Alfred Mann (U. Pennsylvania) and Eugene 
Terray (Woods Hole),
for their constructive and interesting inputs.  I thank 
Joseph Keller (Stanford)
for his time in answering my queries about the geometrical theory of diffraction
and physical optics.

\end{document}